\documentclass[journal=jacsat,manuscript=article]{achemso}
\usepackage{chemformula} % Formula subscripts using \ch{}
\usepackage[T1]{fontenc} % Use modern font encodings

\usepackage{graphicx, amsmath, amssymb}% Include figure files
\usepackage{dcolumn}% Align table columns on decimal point
\usepackage{bm}% bold math
\usepackage{siunitx}
\usepackage{color,soul}
\usepackage[bookmarks,bookmarksopen,pdfstartview=FitH]{hyperref}
\hypersetup{colorlinks=true, urlcolor=blue, linkcolor=red, citecolor=blue}

\author{Wengang Zhang}\email{wzhang01@wesleyan.edu}
\affiliation{Materials Science and Engineering Division, National
    Institute of Standards and Technology, Gaithersburg, Maryland 20899}
\affiliation{Department of Physics, Wesleyan University, Middletown,	Connecticut
    06459-0155}

\author{Francis W.~Starr} \email{fstarr@wesleyan.edu}
\affiliation{Department of Physics, Wesleyan University, Middletown,
    Connecticut 06459-0155}

\author{Jack F.~Douglas}\email{jack.douglas@nist.gov}
\affiliation{Materials Science and Engineering Division, National
    Institute of Standards and Technology, Gaithersburg, Maryland 20899}

\title{Collective Motion in the Interfacial and Interior Regions of Supported
    Polymer Films and its Relation to Relaxation}
\begin{document}

\begin{abstract}
To understand the role of collective motion in the often large changes in
interfacial molecular mobility observed in polymer films, we investigate the
extent of collective motion in the interfacial regions of a thin supported
polymer film and within the film interior by molecular dynamics simulation.
Contrary to commonly stated expectations, we find that the extent of collective
motion, as quantified by string-like molecular exchange motion, is similar in
magnitude in the polymer-air interfacial layer as the film interior, and
distinct from the bulk material. This finding is consistent with Adam-Gibbs
description of the segmental dynamics within mesoscopic film regions where the
extent of collective motion is related to the configurational entropy of the
film as \textit{whole} rather than a locally defined extent of collective motion
or configurational entropy.
\end{abstract}
    
    \maketitle
    
It is generally appreciated that thin supported polymer films, and other
polymeric nanoconfined materials (nanocomposites, spherical polymer
nanoparticles, polymer nanotubes, etc.), exhibit large gradients of mobility in
their interfacial regions that can greatly influence their end-use
properties~\cite{Sharp2003, Ediger2014d, DeMaggio1997, Priestley2005,
    hesami2018molecular}.  Typically, depending on the type of interface and the
nature and magnitude of the interaction strength and the material properties of
the surrounding medium, the scale of the interfacial
regions~\cite{Napolitano2017} with altered mobility is on the order of a few nm,
and the relaxation time in the interfacial region of glass-forming materials can
differ from the overall relaxation time of the film by a
factor~\cite{zhu2011surface, Paeng2012, brian2013surface, zhang2017decoupling,
    bell2003nanometer} as large as $10^{7}$. Since mobility gradients in thin films
can evidently be quite large, it is not surprising that this phenomenon has
elicited significant research interest from both theoretical and practical
perspectives.
    
Changes in mobility of this magnitude can be rationalized within a widely
utilized framework for understanding the slowing down of relaxation in bulk
glass-forming liquids introduced by Adam and Gibbs (AG)~\cite{Adam1965}.
Specifically, a dramatic enhancement of mobility might be interpreted in terms
of a reduction of the scale of collective motion in thin films, and some
evidence for a reduced degree of collective motion has been reported based on
molecular dynamics simulations~\cite{shavit2013evolution,
    riggleman2006influence}. However, this former work did not consider how
collective motion is altered in the interfacial region, but only for the film as
a whole. In particular, Riggleman et al.~\cite{shavit2014physical} observed the
scale of collective motion in thin polymer films to decrease somewhat as the
films were made thinner, a trend notably contrary to what one might naively
expect from the Gibbs-DiMarzio model of glass-formation~\cite{gibbs1958nature}
where a reduction of system dimensionality should lead to a reduction of the
configurational entropy $S_c$ of fluid~\cite{xu2016entropy}, and a corresponding
increase of the glass transition temperature. The fact that many experimental
studies indicate an apparent depression of the glass transition temperature
$T_g$ is often taken as a point against the configurational entropy description
of glass-formation~\cite{jackson1991glass}. However, this criticism does not
apply to the Adam-Gibbs model where structural relaxation time depends both on
the activation free energy in the high temperature fluids ($\Delta \mu$) and
$S_c$. We shall see below that $\Delta \mu$ plays a central role in
understanding the dynamics of our thin films.
    
The identification of large structural relaxation times with a relatively high
degree of collective motion has often been taken to imply that collective motion
in the polymer-air interfacial region should be greatly suppressed with respect
to the interior of the film~\cite{salez2015cooperative, capaccioli2012mechanism,
    ngai1998reduction, ellison2004dramatic}, and Forest and
coworkers~\cite{salez2015cooperative, arutkin2016cooperative} have recently
introduced a model of the interfacial dynamics of glassy materials based on a
combination of the Adam-Gibbs model and the free volume
model~\cite{williams1955temperature} of glass-formation, in which a direct
relation between local density and mobility is postulated. A number of authors
have reported relaxation in the polymer-air interfacial region to be more nearly
Arrhenius than the relaxation of the film as a whole~\cite{yang2010glass,
    zhang2017decoupling, Zhang2017}, seemingly supporting this interpretation of
high interfacial mobility.  However, this attractive interpretation of the
mobility gradient in the interfacial regions of polymer materials raises a
fundamental question with regard to the AG model of glass-formation, since the
scale of collective motion (defined by the number of molecules involved in
cooperative rearrangement) is predicted to scale inversely with the
configurational entropy, arguably a property of the film as a whole (as opposed
to being defined locally). We may then expect that the scale of collective
motion to be the \textit{same} in the film interfacial regions and the film
interior.
    
Based on these considerations, it is a matter of theoretical and practical
interest to quantify how cooperative motion, identified in many earlier works as
string-like particle exchange motion~\cite{donati1998stringlike,
    aichele2003polymer, betancourt2015quantitative, betancourt2013fragility,
    zhang2015role, hanakata2014interfacial, zhang2013string}, varies in the
interfacial and interior regions of model thin supported polymer films.  We
consider a range of film thicknesses and polymer substrate interaction strengths
to evaluate the extent to which AG ideas apply in highly confined materials. We
find that, while the average string length $L$ is reduced relative to the bulk
material,  $L$ varies only weakly when averaged over the interfacial region
compared to that of the film interior.
    
Thus, the large mobility gradients in the film profile are not accompanied by a
corresponding variation in cooperative motion. Evidently, the large mobility
gradient in the interfacial regions arises from the spatial variation of the
\textit{activation enthalpy and entropy within the film}, an effect that
persists even at elevated temperature~\cite{hesami2018molecular}, and which
depends on the boundary interaction strength and film thickness. These
activation free energy parameters then exert a significant influence on changes
of the dynamics observed in thin films, as observed in an earlier
study~\cite{Hanakata2015}.
    
Since the idea of a gradient in the extent of collective motion near interfaces
is an intuitively attractive concept, we also explore the degree of collective
motion layer by layer by binning the strings according to their center of mass
positions normal to the interface to define a ``local'' measure of the extent of
collective motion in the inset of Fig.~\ref{fig:L-vs-T}. Unfortunately, this
measure of local collective motion does not seem to inform about layer by layer
variations of segmental mobility (see inset of Fig.~\ref{fig:L-vs-t}). This
finding is reminiscent of our previous observation~\cite{starr2002we} that the
local density, as defined by local Voronoi volume neighborhoods, is also not
predictive of local molecular mobility. Moreover, previous work has also shown
that the gradient of the local density in the interfacial region of films does
not correlate strongly with the interfacial mobility
gradient~\cite{Hanakata2012, Hanakata2015}.
%%%%%%%%%%%%%%%%%%%%%%%%%%%%%%%%%%%%%%%%%%%%%%%%%%%%%%%%
    
\section*{Results and Discussion}

\begin{figure}[ht!]
        \centering
\includegraphics[width=.7\linewidth]{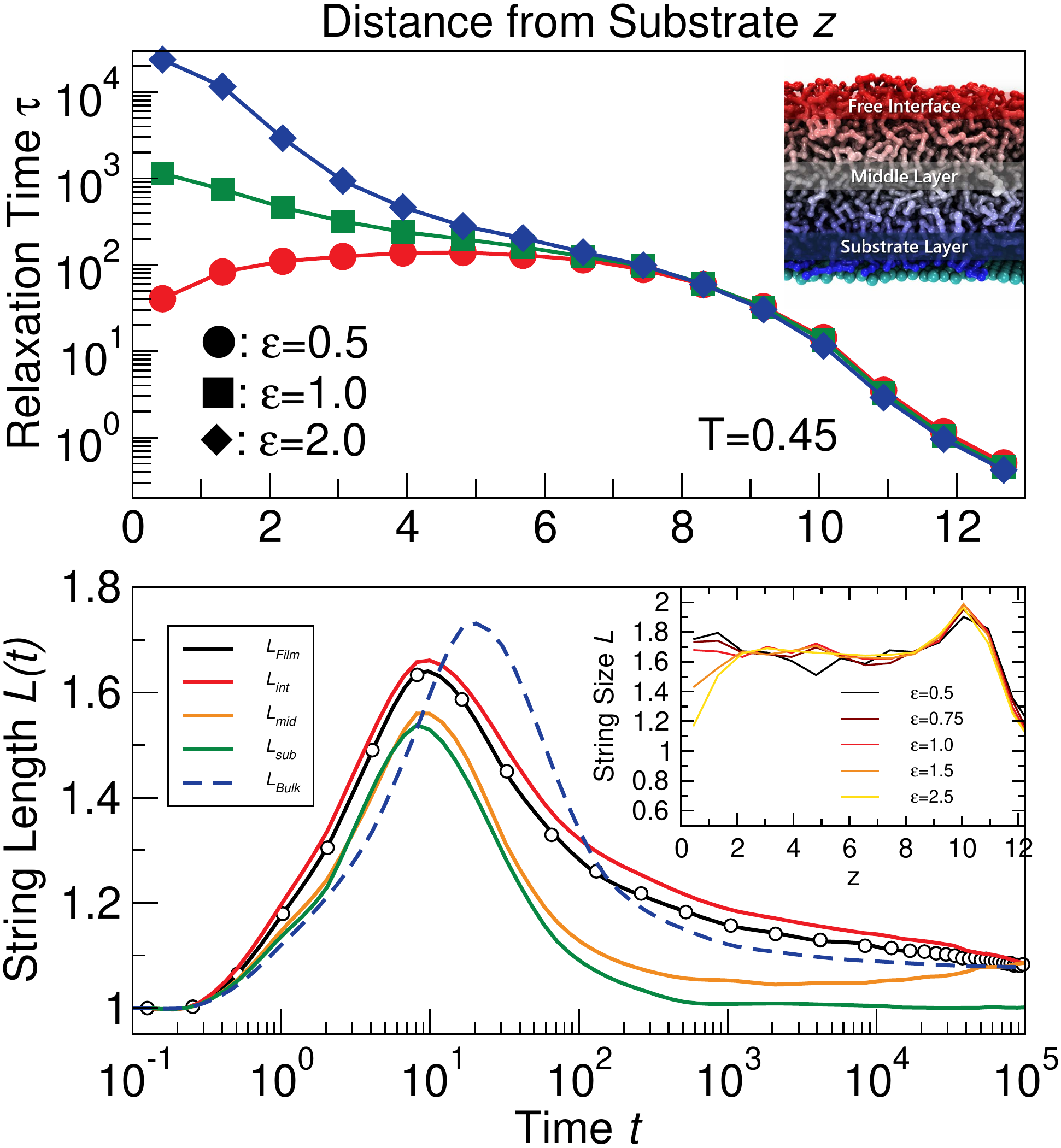} %
% \captionsetup{justification=raggedright,singlelinecheck=false}
\caption{(a) The relaxation profile of polymer film with some
    representative polymer-substrate interaction strength $\varepsilon$
    values at $T=0.45$.  The thin polymer film in the picture has a
    thickness of roughly $12$~nm with a strong substrate-polymer interaction
    strength $\varepsilon = 2.0$. The picture also shows the schematic
    definition of free interface, middle, substrate layer, and substrate
    (turquoise).  (b) The dynamical string length $L(t)$ for free interface,
    middle and substrate layers of the film, and film as a whole.  The
    characteristic peak denotes time scale or ``string
    lifetime''~\cite{starr2013relationship} and the peak string size,
    defining the string length $L$ are similar for different parts of the
    film, despite the significant relaxation gradient within the film. The
    dashed line shows $L$ for the bulk polymer under pressure $P=0$. The
    inset shows the string size as a function of its center of mass
    position. It shows that the variation of string size is relatively
    small, and the string size is nearly the same when averaged over the
    interfacial regions. } %
\label{fig:L-vs-t}
    \end{figure}
    
Our findings are based on an analysis of molecular dynamics simulations of thin
polymer films with variable polymer-substrate interaction strength $\varepsilon$
and film thickness $h$. These simulations have been described in earlier
work~\cite{Zhang2017-2, Zhang2018}.  In brief, we study simulated supported
films composed of a collection of coarse-grained polymers. These polymer films
consist of 320, 400, 480, or 600 polymer chains with film thicknesses $h \approx
8\, \sigma$, $10\, \sigma$, $12\,\sigma$, and $15\,\sigma$, respectively. We
refer to these film as $h=8$, 10, 12, 15 for simplicity. Details of our modeling
and simulations are described in the Simulation Methods section. The reduced
Lennard Jones units can be mapped onto physical units, such as for polystyrene,
by taking $\sigma\approx $ 1 to 2 nm, 1 time unit $\approx $ 9 to 18 ps, and
$\varepsilon/k_{\rm B} \approx 490\,{\rm K}$~\cite{kroger2004simple}.
    
In previous studies on polymer thin films and nanocomposites, we focused on the
relation between the degree of collective motion within the material and the
structural relaxation time, as estimated from the intermediate scattering
function.  In particular, the $T$ dependence of the activation free energy from
relaxation time was determined from simulation, and this quantity was found to
be consistent with the extent of collective motion in the form of string-like
collective segmental exchange events~\cite{Hanakata2015,
    betancourt2015quantitative}, much as Adam-Gibbs has argued for intuitively in
their theory of glass-formation~\cite{Adam1965}. We have found that this
``string model'' of glass-formation \textit{quantitatively} describes relaxation
over the computationally accessible temperature range for a broad range of
systems (bulk polymers, thin supported films, and nanocomposites), as well as
variable material conditions [fixed pressure~\cite{xu2017influence,
    xu2016stringlike}, constant volume~\cite{xu2016influence}, and variable cohesive
interaction~\cite{xu2014influence}]. The present work extends this approach to
consider relaxation in \textit{local} regions within a model glass-forming
liquid.

We first examine the string-like collective motion $L(t)$ in thin polymer films
as a function of both substrate interaction strength $\varepsilon$ and thickness
$h$, following procedures developed in earlier works~\cite{donati1998stringlike,
    aichele2003polymer}. It has been shown that the string-like cooperative motion
is a candidate to quantify cooperatively rearranging regions (CRRs), which
follows the growth of the relaxation activation
energy~\cite{starr2013relationship} and the average length of the strings $L$,
defined by its peak value (See Fig.~\ref{fig:L-vs-t}b) has been found to scale
inversely to the configurational entropy to a good
approximation~\cite{starr2013relationship}, consistent with basic assumptions
made in the Adam-Gibbs model of glass formation when $L$ is equated to the
hypothetical CRR of this model. We consider the string length $L(t)$ for the
film as a whole as well as in the free interfacial region, middle region, and
substrate region, as shown in Fig.~\ref{fig:L-vs-t}.  From the approach
described in Ref.~\citenum{Zhang2018}, the thickness of the substrate layer
$h_{\rm sub}$ nearly saturates for film thickness $h \gtrsim 8$. For simplicity,
we choose substrate layer thickness $h_{\rm sub} = 4.17$, or $\approx 4$ nm in
physical units for the range of film thickness in this study.  In the case where
there is no bound layer~\cite{Zhang2017-2} near the substrate ($\varepsilon
<1.0$), we use the same value for $h_{\rm sub}$, so that we can have a
comparable scale to define the substrate layer relaxation and string length. The
thickness of the ``free'' or polymer-air interfacial layer is defined by the top
part of the film having a thickness of $3.5\,\sigma$, corresponding to 3.5 to 7
nm.  This layer has nearly the same relaxation time for films with different
thickness $h$ and polymer-substrate interaction strength $\varepsilon$.  The
middle layer is defined by the remaining part of the film (i.e.\ the film
excluding the free interfacial and substrate layers). To examine the average
string length $L$ in different regions of the film, we first identify the
strings from the whole film, and sort these strings spatially based on the
position of the center of mass of each string. As illustrated in
Fig.~\ref{fig:L-vs-t} the cooperative motion scale $L(t)$ and the timescale
($t_L$) at which string length peaks in each region, are nearly the same for the
free interface, middle and substrate layers, in spite of differences in the
ratio of local relaxation time between substrate layer and free interfacial
layer being as large as $10^5$.  This observation is consistent with the notion
that the thermodynamic CRR size is not a locally defined quantity. We note that
in previous work based on the present polymer model in the bulk it was shown
that $L$ scales inversely proportional to $S_c$ to a good approximation over the
computationally accessible $T$ range~\cite{starr2013relationship}. %
    
    \begin{figure}[h]
        \centering
\includegraphics[width=.7\linewidth]{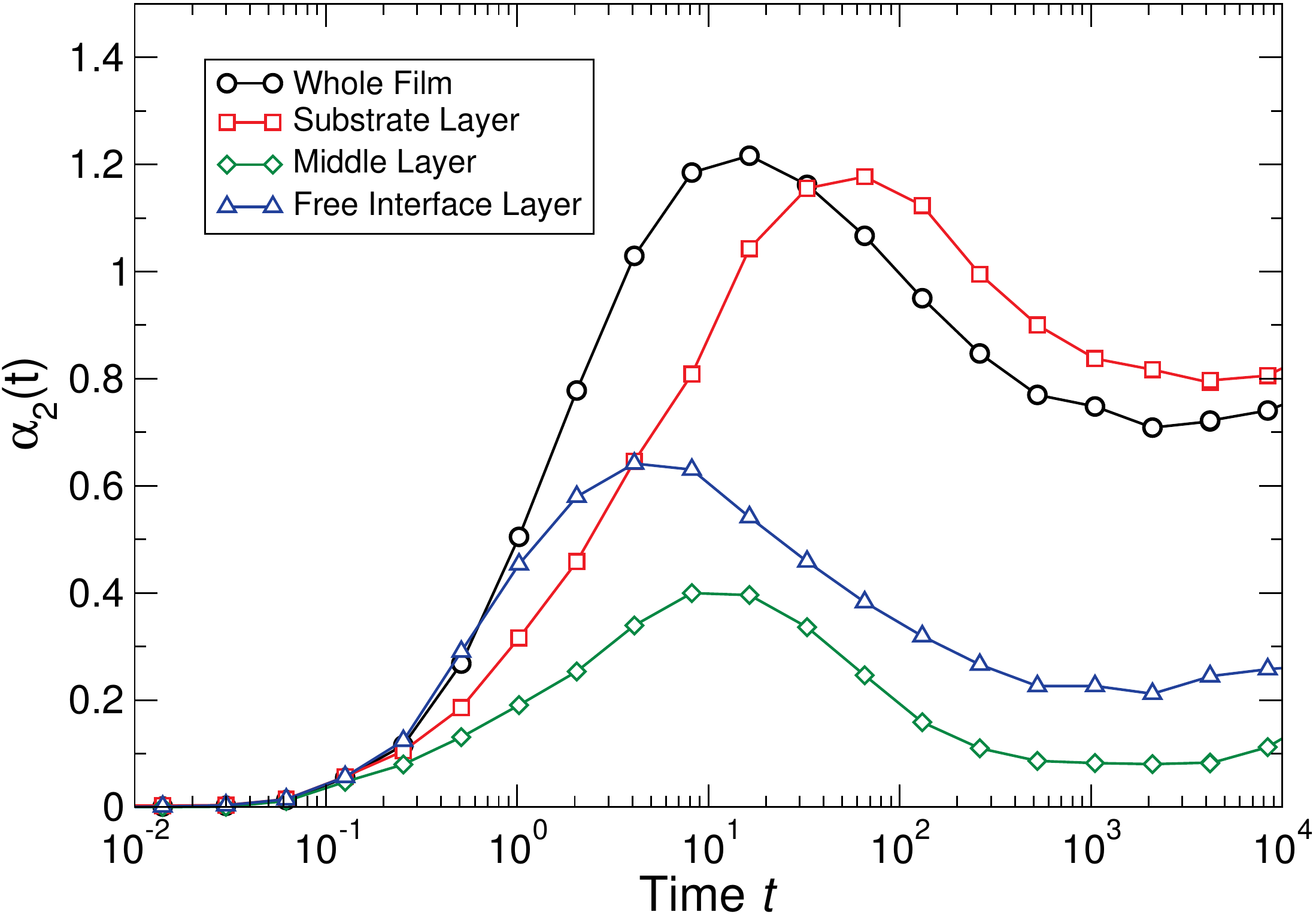} % 
% \captionsetup{justification=raggedright,singlelinecheck=false}
\caption{The non-Gaussian Parameter $\alpha_2(t)$ for free interface,
    middle and substrate layers of the film, and film as a whole at $T=0.5$.
    The thin polymer film has thickness roughly equals to $12\,\sigma$ with
    a strong substrate-polymer interaction strength $\varepsilon = 2.0$. We
    use the same film and layer definition as that in Fig.~\ref{fig:L-vs-t}.
} \label{fig:alpha2}
    \end{figure}
    
It is important to clarify that the near uniformity of the scale of collective
motion in our supported polymer films \textit{does not imply} that ``dynamical
heterogeneity'' within the film is uniform. We support this statement by
considering a common metric of dynamical heterogeneity, the non-Gaussian
parameter,
\begin{equation}\alpha_2(t) = \frac{3\left\langle \Delta \boldsymbol r^{4}(t)
  \right\rangle}{5\left\langle \Delta \boldsymbol r^{2}(t)
        \right\rangle^2}-1,\end{equation} where
    $\left\langle \Delta \boldsymbol r^{2}(t) \right\rangle$ is the mean-square
displacement of the monomers.  This quantity peaks at a characteristic time
$t^*$, related to diffusion in small molecular
liquids~\cite{starr2013relationship}, and defines a segmental mobility time
scale for polymers~\cite{xu2017influence}. In general, $t^*$ exhibits a
power-law scaling in relation to the segmental relaxation time
$\tau_\alpha$, i.e., $t^* \propto \tau_{\alpha}^\xi$, where $\xi< 1$, a
phenomenon termed ``decoupling''~\cite{Hanakata2015, Henritzi2015}. As
expected from the pronounced gradient of mobility, $t^*$ and the height of
the non-Gaussian parameter both vary strongly with their location in the
film in Fig.~\ref{fig:alpha2}. Note that $\alpha_2(t)$ does not vanish at
large $t$ for the film as a whole, or in the interfacial regions, owing to
the gradient in mobility that persists over all time
scales~\cite{Zhang2017}.

From an Adam-Gibbs perspective, collective molecular motion is important for
understanding the structural relaxation in glass-forming systems. Naively, the
apparent invariance of string size to location in the film would lead us to
expect that the AG picture cannot be extended to understand the extreme
variations in local relaxation. However, as we now discuss, the physical
situation is more subtle. To apply the AG approach \textit{locally}, we examine
the dynamics of each film region using string model of relaxation in
glass-forming materials~\cite{pazmino2014string, betancourt2015quantitative}, a
modern extension of the Adam-Gibbs model founded on simulation evidence,
    \begin{equation}
    \tau (T) =  \tau_0(\varepsilon, h) \exp\left[ \frac{L(T)}{L_{\rm A}}
    \frac{\Delta \mu(T)}{k_{\rm B}T} \right], 
    \label{eq:string-model}
    \end{equation} 
where $\tau_0 = \tau_{\beta}(\varepsilon, h) \exp\left[ \frac{-\Delta \mu(T_{\rm
        A})}{k_{\rm B}T_{\rm A}} \right]$ with $\tau_{\beta} \equiv \tau(T_{\rm A})$ and
$\Delta \mu(T, \varepsilon, h) = \Delta H (\varepsilon, h)- T\Delta
S(\varepsilon, h)$; $T_{\rm A}$ is the onset temperature of glass
formation~\cite{betancourt2015quantitative, starr2013relationship}, and $\Delta
H$ and $\Delta S$ are the high temperature enthalpic and entropic contributions
of the free activation energy, respectively; $\tau_{\beta}$ is the fast $\beta$
relaxation time, which equals the $\alpha$-relaxation time $\tau_{\alpha}$ at
$T_{\rm A}$~\cite{pazmino2018string}.  In the bulk material, $\Delta H$ is
directly related to the activation energy $E_a$ determined from fitting
relaxation time over high temperature region where relaxation is
Arrhenius~\cite{xu2016influence}. We utilize a fixed onset temperature $T_{\rm
    A} = 0.65$ for thin films, as estimated in Ref.~\citenum{pazmino2014string},
since its value is relatively insensitive to polymer film thickness and
polymer-substrate interaction strength. $L_{\rm A} \equiv L(T_{\rm A})$ is the
string length at the onset temperature $T_{\rm A}$, the residual collective
motion in the high temperature liquid~\cite{Hanakata2015}. Note that both
$L_{\rm A}$ and $\tau_{\rm A}$ depend on film thickness $h$, polymer-substrate
interaction strength $\varepsilon$, as well as in the different regions of the
film with, but the range of value is not large, $L_{\rm A} = 1.40 \pm 0.02$ and
$\tau_{\rm A} = 2.3 \pm 1.0$~\cite{betancourt2015quantitative}. We emphasize
that $\tau_{\rm 0}$ is not a free fitting parameter, but $\tau_0$ rather is
determined ~\cite{Hanakata2015} by $\Delta H$ and $\Delta S$.  It is also
notable that $\tau_{0}$ varies significantly with film thickness $h$, along with
the supporting boundary interaction strength and stiffness~\cite{Hanakata2015}.

\begin{figure}[h]
\centering \includegraphics[width=.8\linewidth]{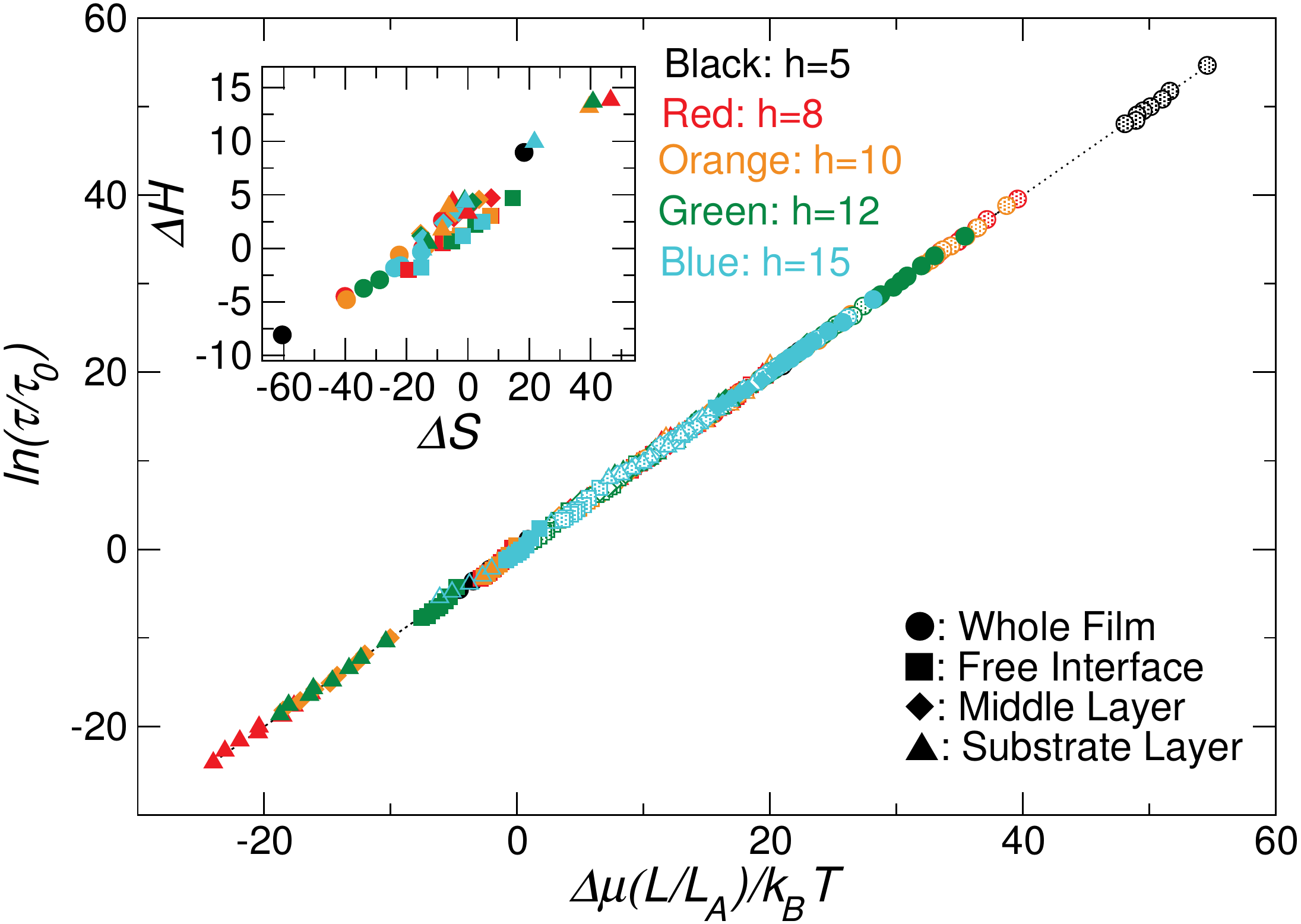} %
% \captionsetup{justification=raggedright, singlelinecheck=false}
\caption{Test of string model for cooperative relaxation. Reduced relaxation
    time as a function of the reduced activation free energy for different film
    thicknesses, and regions of the film (free interfacial, middle and substrate
    layer). Hollow, dotted, and solid symbols stand for $\varepsilon = $ 0.5, 1.0,
    and 2.0, respectively. The inset shows the entropy-enthalpy compensation plot,
    obtained from fit to Eq.~\ref{eq:string-model}.}%
 \label{fig:string-model-plot}
    \end{figure}
    
Although the approximate invariance of string size to location within the film
does not explain the large variations in relaxation time within films, we may
use Eq.~\ref{eq:string-model} to understand the dynamics of thin polymer film
for both film as a whole and local regions within it, and the relation of film
relaxation to that of the bulk material. Figure~\ref{fig:string-model-plot}
shows that there is a linear relationship between the reduced relaxation time
$\ln \left( \tau/\tau_0\right)$ and the reduced activation energy
$\frac{L(T)}{L_{\rm A}} \frac {\Delta \mu} {k_{\rm B}T}$ for various
polymer-substrate interaction strength $\varepsilon$, film thickness $h$, and
different local regions. A remarkable universal collapse of $\tau$ in term of
string length was found in Refs.~\citenum{hanakata2014interfacial, Hanakata2015,
    betancourt2015quantitative} for both thin polymer film and polymer
nanocomposites. Note that in the case of the extremely thin film with thickness
$h=5$, the free interfacial and substrate layer are not well-defined, so
Fig.~\ref{fig:string-model-plot} does not include these interfacial regions. We
thus find the string model of glass formation~\cite{pazmino2014string} can also
quantitatively describe \textit{local} film dynamics.  The values of $\Delta H$
and $\Delta S$ (fitting parameters) that result from the application of
Eq.~\ref{eq:string-model} are shown in the inset of
Fig.~\ref{fig:string-model-plot}, which, when plotted parametrically, show that
an entropy-enthalpy compensation relation ($\Delta S= \Delta S_0 + T_{\rm comp}
\Delta H$, where $T_{\rm comp} \approx 0.21$) holds for different film regions
as well. The value of $T_{\rm comp}$ obtained here is consistent with a previous
estimate obtained from thin film and polymer nanocomposites simulations based on
the same polymer model~\cite{betancourt2015quantitative}.

Large gradients and an entropy-enthalpy compensation relation have also been
observed in the interfacial dynamics of crystalline Ni~\cite{zhang2015influence}
and Cu~\cite{papageorgiou2000multiple} so that this phenomenon apparently arises
in both crystalline and non-crystalline materials.  Note that $\Delta H$ and
$\Delta S$ estimates in the inset of Fig.~\ref{fig:string-model-plot} near the
solid substrate can be negative. This counter-intuitive phenomenon has been
observed in the kinetics of highly confined fluids when the interaction between
the molecule and the boundary are strongly
attractive~\cite{artioli2013catalysis}. Entropy-enthalpy compensation and
negative values of $\Delta H$ and $\Delta S$ are also commonly observed in the
thermodynamics of molecules binding~\cite{dudowicz2018lattice}, a
counter-intuitive phenomenon associated with competitive molecular interactions.
Previous work investigating the mobility gradient near the free interface of a
crystalline material (Cu) ~\cite{papageorgiou2000multiple}, has quantified the
mobility gradient in terms of a gradient in the activation free energy $\Delta
\mu$, and we likewise consider the segmental relaxation time $\tau_\alpha$ and
activation free energy as a function of distance from the film free surface. In
particular, if we take $z=0$ to denote the position of the polymer interface and
$L$ to be the average value of the film as whole, then the string model
prediction for the segmental relaxation time can be formally written,
    \begin{equation}
    \frac{\tau_{\alpha}(z)}{\tau_{0}(z)} = \left(\frac{\tau_{\alpha}^{\rm
            Film}}{\tau_{0}^{\rm Film}}\right)^{\frac{\Delta \mu(z)}{\Delta \mu ^{\rm
                Film}}},
    \label{eq:power-law}
    \end{equation}
    where $\tau_{\alpha}^{\rm Film} = \tau_0^{\rm Film} \exp\left[
\frac{L(T)}{L_{\rm A}} \frac {\Delta \mu^{\rm Film}} {k_{\rm B}T} \right]$
is the relaxation time for the film as a whole. The large gradient in the
relaxation time $\tau_\alpha$ within the polymer film within this model can
thus be traced to a gradient in $\Delta \mu$ rather than a variation of the
extent of collective motion as a function of distance within the film.
Moreover, by averaging over interior and interfacial regions of the film, we
obtain an extension of Eq.~\ref{eq:power-law} that relates the ratio of
interior and polymer-air interfacial relaxation times to the difference in
the mean activation free energy in these regions, namely we have the
relaxation time ratio,
    \begin{equation}
    \frac{\tau_{\rm mid}}{\tau_{\rm int}} = \frac{\tau_0^{\rm mid}}{\tau_0^{\rm
            int}} \exp \left[\frac{L(T)}{L_{\rm A}} \frac{(\Delta \mu_{\rm mid} - \Delta
        \mu_{\rm int})}{k_{\rm B}T} \right],
    \label{eq:tau-ratio}
    \end{equation}
where $L(T)$ is again the characteristic string length of the film as a
\textit{whole}. This relation is potentially of significant practical value
since the ratio $\tau_{\rm mid}/\tau_{\rm int}$ is experimentally accessible,
and recent measurements have indicated that this mobility ratio can be as large
as $10^{7}$ near $T_g$~\cite{zhang2017decoupling, brian2013surface}.  Note that
Eq.~\ref{eq:tau-ratio} can be well-approximated as a Vogel-Fulcher-Tammann (VFT)
function~\cite{williams1955temperature} over the computationally accessible
temperature range, and we previously found this ratio to extrapolate to a value
on the order of $\mathcal{O}(10^{11})$ as $T$ approaches $T_g$~\cite{Zhang2017}.

\begin{figure}[h]
 \centering \includegraphics[width=.7\linewidth]{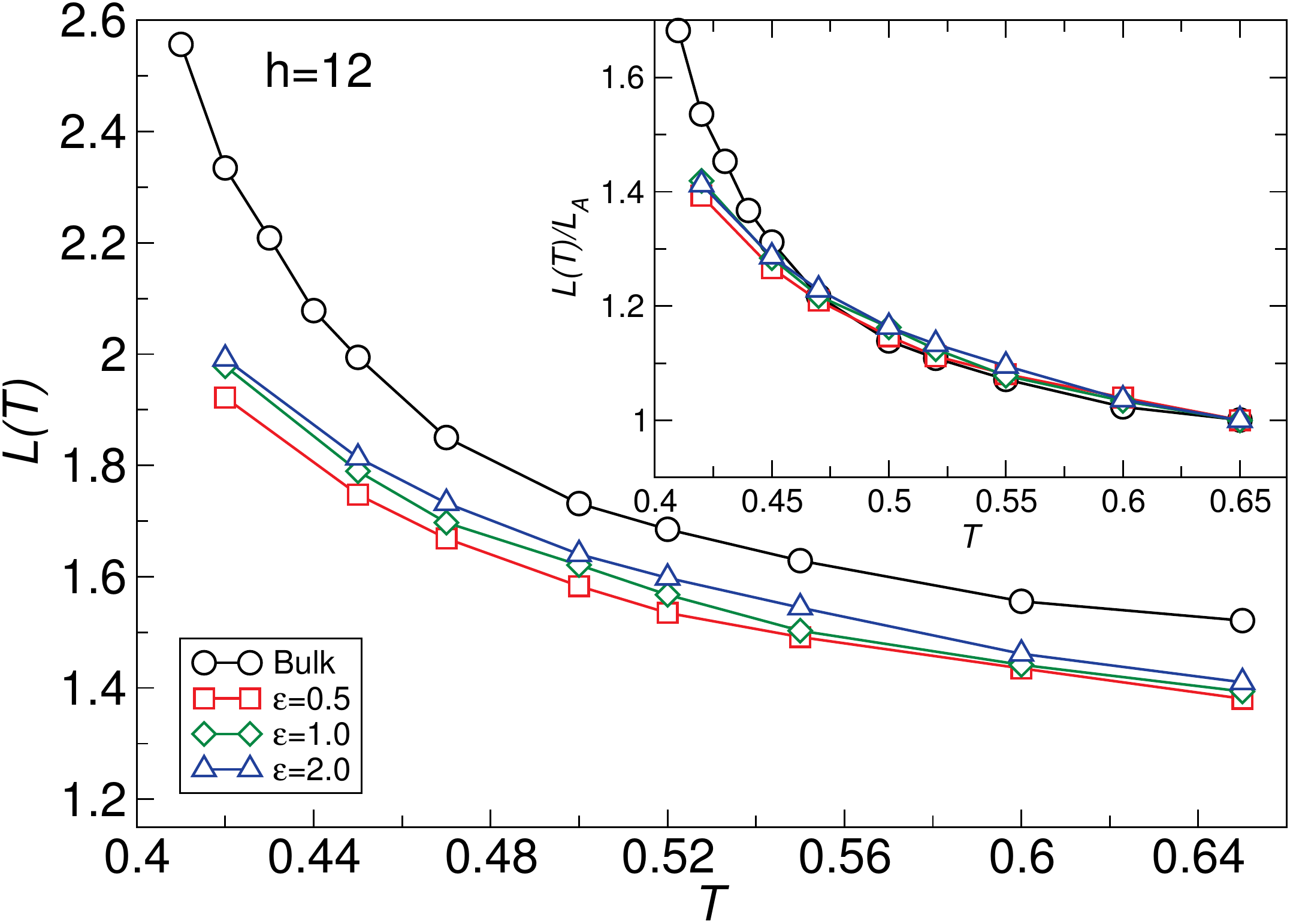}%
\caption{The characteristic string length $L(T)$ for polymer films as a whole
    with some representative polymer-substrate interaction strengths ($\varepsilon
    =$ 0.5, 1.0, and 2.0), and bulk polymer.  The inset shows $L(T)$ normalizes by
    their corresponding string length at the onset temperature $L_{\rm A}$ ($L_{\rm
        A} \equiv L(T_{\rm A})$).  Although confinement and polymer-surface interaction
    strength influence the scale of collective motion, the relative change in $L$
    is rather insensitive to their confinement and interfacial variables. }
\label{fig:L-vs-T}
    \end{figure}

While the extent of collective motion clearly changes with film thickness, we
may still approximately relate relaxation within the film to relaxation of the
bulk material. Provided the \textit{ratio} $L/L_{\rm A}$ remains nearly the same
in thin film and bulk material with $L_{\rm Bulk}/L_{\rm A}^{\rm Bulk} \approx
L_{\rm Film}/L_{\rm A}^{\rm Film}$ (see Fig.~\ref{fig:L-vs-T}), we then have the
approximate relation:
    \begin{equation}
    \begin{aligned}
    \frac{\tau_\alpha(z)}{\tau_{0}(z)} &\approx
    \left(\frac{\tau_{\alpha}^{\rm Bulk}}{\tau_{0}^{\rm
            Bulk}}\right)^{\frac{\Delta \mu(z)}{\Delta \mu^{\rm Bulk}}}.
    \end{aligned}
    \label{eq:power-law-2}
    \end{equation}
Figure~\ref{fig:L-vs-T} shows that the relative change in collective motion
$L/L_{\rm A}$ is indeed similar in magnitude in the bulk and thin film for the
temperature range and polymer-substrate interaction strength range. We emphasize
that Eq.~\ref{eq:power-law-2} is only suggested to be a reasonable
\textit{approximation} over the computationally accessible temperature range.
Nonetheless, Eqs.~\ref{eq:tau-ratio} and~\ref{eq:power-law-2} allow for an
alternate understanding of previously observed computational evidence for a
phenomenological power law or ``decoupling'' relation linking the relaxation
time of the film as a whole to the relaxation times within the interfacial
regimes and between the film as a whole and the bulk
material~\cite{phan2018dynamic, diaz2018temperature, chung2017wrinkling}.  The
near constancy of $L/L_{\rm A}$ between the bulk and thin films, along with the
normally reduced molecular cohesive interaction strength at the polymer-air
boundary, also naturally explains the near-Arrhenius relaxation in the
interfacial region, its relatively high mobility in comparison to the bulk. It
will be interesting to see whether the ``decoupling'' relation between
wave-vector dependence of the relaxation time $\tau(q)$ from the intermediate
scattering function to $\tau_\alpha$ can likewise be understood in a similar
fashion since the scale of collective motion must also be independent of
observational scale.

%%%%%%%%%%%%%%%%%%%%%%%%%%%%%%%%%
\section*{Conclusions}

Our investigation of collective motion in relation to the internal dynamics of
thin supported film provides further evidence of the importance of variations of
the cohesive interaction in thin films for understanding both changes in
relaxation in relation to the bulk and mobility gradients with these films.
Changes in the cohesive interaction strength in the interfacial region are
important because they alter the activation free energy $\Delta \mu$, which
effects even the liquid regime far above the glass transition
temperature~\cite{Hanakata2015, betancourt2015quantitative}. We find that many
important aspects of the dynamics can be understood from interfacial changes of
$\Delta \mu$, rather than changes in the scale of collective motion.
Specifically, we find (i) a greatly enhanced mobility at the polymer-air
interface in comparison to the bulk material, (ii) a near invariance of enhanced
interfacial mobility with changes of film thickness, and (iii) the phenomenon of
enthalpy-entropy compensation in the activation free energy parameters, $\Delta
H$ and $\Delta S$. The observation of a similarity in the degree of cooperative
motion within the polymer-air interfacial region and the film interior is also
theoretically interesting because it provides guidance regarding how to extend
the Adam-Gibbs model to describe local mobilities within mesoscopic regions in
glass-forming polymer films, and potentially nanocomposites.  We plan to extend
the analysis of the present paper to describe nanocomposites and to understand
the physical origin of the width of the interfacial zones on the scale of
collective motion found previously for both polymer films and nanocomposites.
    
An important practical implication of our work is the suggestion that local
changes in the segmental dynamics can be understood primarily from changes
dynamics of the Arrhenius activation free energy parameters characteristic of
the fluid dynamics at elevated temperatures.  Perhaps surprisingly, the extent
of collective motion within the  film does not vary substantially across the
film profile, and accordingly does not contribute significantly to spatial
variations in the local dynamics. Consequently, knowing the extent of collective
motion for the material as a whole appears to be sufficient to understand change
of material dynamics with confinement, if the changes to local activation
parameters are additionally known. This is good news, since, if one had to
instead determine the degree of collective motion locally to understand local
mobility variations, then the theory would be essentially intractable from a
practical standpoint. The deduction of simple power-law relations between the
segmental relaxation of the film as a whole only exists in the string model of
glass-formation when the string length parameter of this model has no depth
dependence. This is apparently the origin of both the decoupling relation in the
string model of glass formation and the success of this model in fitting our
simulation data for the relaxation time as function of depth over a wide range
of temperatures. The fact that the spatial dependence of the cooperativity scale
is not needed provides a readily implemented framework for studying mobility
variations in glassy materials.

Of course, the general validity  of this extended string model of
glass-formation requires further confirmation in polymer nanocomposites and
other types of non-uniform glass-forming materials to test the validity of this
model. There is also a need to better understand the root physical causes of the
gradients in the activation energy parameters given their large influence on the
mobility gradients in thin films. In bulk materials, the high temperature
activation enthalpy  correlates very strongly with the cohesive energy density
of the fluid, suggesting that these gradients in activation energy may derive
physically from gradients in the potential energy density near the interfaces,
which we are currently investigating. Further efforts are also required to
understand the ubiquitous enthalpy-entropy compensation relation linking the
activation enthalpy to activation entropy in the dynamics of many condensed
materials. We suggest that more theoretical and experimental efforts should be
devoted to understanding these fundamental energetic parameters.

\section{Modeling and Simulation Details}%
Our results are based on molecular dynamics simulations of thin supported
polymer films.  We simulate supported thin polymer films with variable film
thickness $h$ and strength of attractive interaction $\varepsilon$ between the
substrate and polymer employing molecular dynamics simulations. The polymer film
model is the same as that used in Refs.~\citenum{Hanakata2015, Zhang2017-2}. The
polymer films have 320, 400, 480, or 600 polymer chains; these films have
thicknesses $h \approx$ $5\, \sigma$, $8\, \sigma$, $10\, \sigma$, $12\, \sigma$
and $15\,\sigma$, respectively, which decrease approaching $T_g$. These films
are referred to as $h=$ 8, 10, 12, and 15. Above the film is free (empty) space,
so the film is effectively at pressure, $P=0$. As a reference for the
thermodynamic and dynamic properties, we also simulated a bulk polymer with
periodic boundary conditions in all directions at pressure $P=0$.
    
Polymers are modeled as unentangled chains of $10$ beads linked by harmonic
springs. We use the harmonic spring potential $U_{\rm bond}=\frac{k_{\rm
        chain}}{2}(r-r_0)^2$ to connect nearest-neighbor monomers within a polymer
chain. The equilibrium bond length is $r_0=0.9$ and the spring constant is
$k_{\rm chain}=1111$ ~\cite{Hanakata2015}. To inhibit crystallization of the
film, we choose $r_0$ smaller than that chosen in
Ref.~\citenum{Peter2006thickness}. We use the same substrate model as that in
Ref.~\citenum{Zhang2017-2} for all the films studied. The substrate consists of
528 particles arranged in a triangular lattice~(the (111) face of an FCC
lattice). We tether substrate particles via a harmonic potential $V_{\rm
    sub}(r)=(k/2)(r-r_0)^2,$ where $r_0$ is the ideal lattice position and $k=50$ is
the spring constant~\cite{baschnagel2005computer, varnik2002shear}. We use
Lennard-Jones (LJ) interactions between non-bonded monomers and substrate
particles. The interactions are truncated at pair separations
$2.5\,\sigma_{ij}$, where $\sigma_{ij}$ is equivalent to the particle diameter
in the LJ potential, and the subscript $ij$ indicates the possible combinations
of interactions~($ss$ substrate-substrate, $ps$ polymer-substrate, $pp$
polymer-polymer). All units are given relative to the strength $\varepsilon$ and
size $\sigma$ of non-bonded polymer-polymer interactions.  Consequently, $T$ is
in unit of $\varepsilon/k_{\rm B}$, where $k_{\rm B}$ is Boltzmann's constant,
pressure is in the unit of $\varepsilon/\sigma^3$, and time in units of $\sqrt{m
    \sigma ^2/\varepsilon}$. The LJ parameters are $\sigma_{pp}=1.0$,
$\varepsilon\equiv\varepsilon_{pp}=1.0$, $\sigma\equiv\sigma_{ps}=1.0$,
$\sigma_{ss}=0.8$, $\varepsilon_{ss}=1.0$, and we use interaction strengths
between monomers and substrate particles $\varepsilon_{ps}=$ $0.1$, $0.25$,
$0.5$, $0.75$, $1.0$, $1.25$, $1.5$, $2.0$, $2.5$, and $3.0$. Since we only vary
$\varepsilon_{ps}$, we simply refer to this quantity as $\varepsilon$.

Periodic boundary conditions are used in the directions parallel to the
substrate with a box length $19.76\,\sigma$ (determined by the lattice spacing
of the triangular lattice substrate).  We conducted all simulations using the
LAMMPS~\cite{Plimpton1995} simulation package with a time step $dt=0.002$. For
cooling and heating simulations of the bulk polymers, we use an {\it NPT}
ensemble at $P=0$. We performed at least 3 independent heating and cooling runs
for both the pure polymer and polymer films at the same rate $10^{-5}$. To
generate trajectories from which we study the dynamics at fixed $T$, we carry
out {\it NPT} simulations starting from configurations taken from the heating
runs at $T>T_g$ with pressure $P=0$. For the supported polymer films, we use an
\textit{NVT} ensemble where the box dimension is the $z$-direction is large
compared to the film thickness. The temperatures are varied from 0.45 to 0.65,
above (the heating rate dependent) $T_g(h=15)\approx 0.40$ of the thickest
polymer film~\cite{Zhang2017-2}. We equilibrate each trajectory for at least 100
times the average polymer relaxation time of the entire film $\tau_{\rm
    overall}$.

%%%%%%%%%%%%%%%%%%%%%%%%%%%%%%%%%%%%%%%%%%%%%%%%%%%%%%
\section*{Acknowledgement} %
Computer time was provided by Wesleyan University. This work was supported in
part by NIST Award No. 70NANB15H282.

% Bibliography
%\bibliography{bib}
 
\providecommand{\latin}[1]{#1}
\providecommand*\mcitethebibliography{\thebibliography}
\csname @ifundefined\endcsname{endmcitethebibliography}
{\let\endmcitethebibliography\endthebibliography}{}

\end{document}